\documentclass[english,aps,pre, reprint, twocolumn]{revtex4}
\usepackage[T1]{fontenc}
\usepackage[latin9]{inputenc}
\usepackage{amsmath}
\usepackage{amssymb}
\usepackage{graphicx}
\usepackage{esint}

\makeatletter
\@ifundefined{textcolor}{}
{%
 \definecolor{BLACK}{gray}{0}
 \definecolor{WHITE}{gray}{1}
 \definecolor{RED}{rgb}{1,0,0}
 \definecolor{GREEN}{rgb}{0,1,0}
 \definecolor{BLUE}{rgb}{0,0,1}
 \definecolor{CYAN}{cmyk}{1,0,0,0}
 \definecolor{MAGENTA}{cmyk}{0,1,0,0}
 \definecolor{YELLOW}{cmyk}{0,0,1,0}
}

\usepackage{bbm}

\makeatother

\usepackage{babel}
\begin{document}

\preprint{\%This line only printed with preprint option}

\title{Dynamics of hydrodynamically coupled Brownian harmonic oscillators
in a Maxwell fluid}

\author{Shuvojit Paul}
\email{sp12ip006@iiserkol.ac.in}

\affiliation{Indian Institute of Science Education and Research Kolkata }
\begin{abstract}
Recently, many interesting features of the hydrodynamically coupled
motions of the Brownian particles in a viscous fluid have been reported
which are impossible for the uncoupled motions of the similar particles.
However, it is expected that those physics in a viscoelastic fluid
is much more interesting due to the presence of the additional frequency
dependent elasticity of the medium. Thus, a theory describing the
equilibrium dynamics of two hydrodynamically coupled Brownian harmonic
oscillators in a viscoelastic Maxwell fluid has been derived which
appears with new and impressive aspects. Initially, the response functions
have been calculated and then the fluctuation-dissipation theorem
has been used to calculate the correlation functions between the coloured
noises present on the concerned particles placed in a Maxwell fluid
due to the thermal motions of the fluid molecules. These correlation
functions appear to be in a linear relationship with the delta-correlated
noises in a viscous fluid. Consequently, this reduces the statistical
description of a simple viscoelastic fluid to the statistical representation
for an extended dynamical system subjected to delta-correlated random
forces. Thereupon, the auto and cross-correlation functions in the
time domain and frequency domain and the mean-square displacement
functions of the particles have been calculated which are perfectly
consistent with their corresponding established forms in a viscous
fluid and emerge with exceptional characteristics.
\end{abstract}
\maketitle

\section{introduction}

The study of the statistical properties of Brownian motion has paramount
importance in Physics with varieties of applications. The random motions
of colloidal particles in a viscous fluid have been studied since
1905 \cite{einstein1905einstein,von1906kinetischen} and thus, well-established
theory exists to describe corresponding statistical properties. Similarly,
the Brownian dynamics of microscopic particles in a viscoelastic fluid
are well investigated \cite{volkov1996non,grimm2011brownian,tassieri2010measuring,mason1995optical},
and these are entirely different from the corresponding dynamics in
a purely viscous fluid. For example, at low Reynold's number approximation
and moreover, in typical experimental time scales, the inertial effect
of a Brownian particle and the vorticity diffusion are negligible.
Consequently, the dynamics of Brownian particles in a viscous fluid
are determined only by the instantaneous forces. Thus, there is no
memory \cite{meiners1999direct} and the process is known as a Markovian
process. But conversely, in a viscoelastic medium, the stochastic
motion of a Brownian particle is non-Markovian, even if the inertia
is negligible. 

Further, the hydrodynamic interaction between the moving particles
in a fluid may exceptionally change the statistical description of
the Brownian motion of individual particles. For instance, two hydrodynamically
coupled particles in a viscous fluid in two time-independent external
potentials can impose time delayed correlations between that two particles
\cite{meiners1999direct,martin2006direct} which is impossible for
a single particle in a viscous fluid with negligible inertial effect.
Again, two viscously coupled oscillators may have a frequency maximum
or motional resonance effect in their mutual response function \cite{paul2017direct}
which is a sensitive function of the fluid viscosity. Thus, this function
can be used for rheological measurements \cite{paul2018two}. Such
study of coupled motions of Brownian particles reveals interesting
physics and definitely helps to understand the dynamics of colloidal
suspensions, microscopic dynamics of proteins, dynamics of polymer
solutions etc. But, as most of the fluids (emulsions, biological fluids,
etc.,) are viscoelastic in nature \cite{brust2013rheology,ayala2016rheological,doi1988theory},
so the study of coupled Brownian dynamics of micro-particles in such
complex fluid have fundamental importance. Understandably, there exists
a strong interest in the scientific community. A viscoelastic fluid
exhibits both viscous and elastic nature and thus it is typically
characterized by a complex frequency dependent viscosity $\eta(\omega)$.
The simplest approach developed to describe linear viscoelasticity
is the Maxwell model. Further, it is improved further into a generalized
form and the corresponding name of the model is Jeffreys' model \cite{doi1988theory,grimm2011brownian}.
It can provide the relation between stress and shear-rate in linear
viscoelastic fluids which can be used to evaluate trajectories of
Brownian particles in such medium. 

Although some statistical properties of the coupled dynamics in a
complex fluid have been used for the purpose of rheological measurements
\cite{crocker2000two,crocker2007multiple}, the coupled dynamics of
Brownian particles in such fluid is still not studied elaborately
which may disclose several interesting physics and can be used to
measure rheological parameters more precisely.

This work theoretically describes the dynamics of two hydrodynamically
coupled Brownian particles bound in two harmonic oscillator potentials
in a Maxwell fluid with single relaxation time. The response functions
of the particles under external perturbations have been calculated
and it has been shown that these functions are entirely different
in a Maxwell fluid as compared to the reported forms in a viscous
fluid \cite{paul2017direct,paul2018two}. Furthermore, the fluctuation-dissipation
theorem (FDT) has been used to calculate the correlations between
stochastic noises on the two particles in the study. It has been shown
that the noise correlations in Maxwell fluid are linearly related
to the correlations of noises in a viscous fluid by a function. Thus,
such problem of coupled Brownian motion with the simplest viscoelastic
liquid can be reduced to the statistical description of an extended
dynamical system subjected to a delta-correlated random force. This
method is reported for a single particle in Maxwell fluid with one
relaxation time \cite{volkov1996non}. Then the position correlation
functions in the frequency domain and in the time domain and the mean-square
displacement functions of the particles have also been calculated
and studied. It has been shown in addition that the correlation between
two particles in Maxwell fluid is time-delayed, but the time delay
linearly depends on the Maxwell relaxation time $\tau$, which represents
the crossover time scale of the fluid from elastic to viscous behaviour.
Consequently, $\tau\rightarrow0$ represent the statistical description
of the particles motion in the viscous fluid. Besides, if the particles
are significantly separated and if the hydrodynamic coupling is negligible
then they behave like uncoupled particles and show corresponding statistical
properties. The zero stiffnesses consideration of the bounding potentials
converges the dynamics to the free but hydrodynamically coupled dynamics
of the concerned particles. 

\section{Theory}

The one-dimensional translational motion of two hydrodynamically coupled
identical colloidal spheres bound in two different harmonic oscillator
potential wells in Maxwell fluid with a single relaxation time $\tau$
can be described by the set of equations \cite{gardiner1984handbook,volkov1996non}
\begin{align}
m\ddot{x}_{1} & =F_{1}-k_{1}x_{1}+\varepsilon(-k_{2}x_{2}+f_{2})+f_{1}\label{eq:1-1}\\
m\ddot{x}_{2} & =F_{2}-k_{2}x_{2}+\varepsilon(-k_{1}x_{1}+f_{1})+f_{2}\label{eq:2-1}
\end{align}

where, $F_{1}$ and $F_{2}$ are the relaxed frictions which are given
by
\begin{align}
\tau\frac{dF_{1}}{dt}+F_{1} & =-\zeta_{0}\dot{x}_{1}\label{eq:3}\\
\tau\frac{dF_{2}}{dt}+F_{2} & =-\zeta_{0}\dot{x}_{2}\label{eq:4}
\end{align}

$x_{1}$, $x_{2}$ are the positions of the first and second particles
respectively w.r.t. their corresponding potential minimums; $k_{1}$
and $k_{2}$ are the stiffnesses of the first and second harmonic
oscillators respectively; $\varepsilon=\frac{3a_{0}}{2d}-\left(\frac{a_{0}}{d}\right)^{3}$
is the hydrodynamic coupling coefficient.; $a_{0}$ is the radius
of each of the particles and $d$ is the center to center separation
between these two particles; $\zeta_{0}=6\pi\eta_{0}a_{0}$ is the
drag coefficient at zero frequency; $\eta_{0}$ is the zero frequency
viscosity of the fluid; $m$ is the mass of each of the particles.
The particles are far apart from each other so that $d$ can be assumed
to be constant in time. $f_{1}$ and $f_{2}$ are the perturbations
on first and second particles respectively, which can be due to the
random motions of the surrounding molecules of the fluid or due to
some external disturbance. In low Reynolds number regime, the effect
of inertia dies out in a time scale $\tau^{*}=m/\zeta_{0}$ which
is known as the momentum relaxation time scale. $\tau^{*}$ is of
the order of $\sim10^{-6}$ sec which is much lower than the typical
experimental time scales. Hence, the inertial effect is negligible
and the terms in the left hand side of Equs. \eqref{eq:1-1} and \eqref{eq:2-1}
can be approximated to zero and thus can be written as

\begin{gather}
0=F_{1}-k_{1}x_{1}+\varepsilon(-k_{2}x_{2}+f_{2})+f_{1}\label{eq:1}\\
0=F_{2}-k_{2}x_{2}+\varepsilon(-k_{1}x_{1}+f_{1})+f_{2}\label{eq:2}
\end{gather}

The systematic viscoelastic drag forces $F_{i}$ ($i=1,\,2$) tend
to $-\zeta_{0}\dot{x}_{i}$ as the relaxation time $\tau$ converges
to zero, which is clear from the set of Equs. \eqref{eq:3},\eqref{eq:4}.
This is the familiar expression of the Stokes drag force in viscous
fluid. Now, the Equs. \eqref{eq:3} and \eqref{eq:4} can be easily
solved to yield the expressions for the drags on the spheres in Maxwell
fluid in terms of the initial forces $F_{i}^{0}$ as 
\begin{equation}
F_{i}(t)=F_{i}^{0}e^{-(t-t_{0})/\tau}-\intop_{t_{0}}^{t}\zeta(t-s)\dot{x}_{i}(s)ds
\end{equation}

where the memory kernel $\zeta(t)=\frac{\zeta_{0}}{\tau}e^{-t/\tau}$.
It can be assumed that the initial time was $-\infty$ and then $F_{i}^{0}=0$.
On this assumption, the expressions for the drags on the spheres are
\begin{equation}
F_{i}(t)=-\frac{\zeta_{0}}{\tau}\intop_{-\infty}^{t}e^{-(t-s)}\dot{x}_{i}(s)ds
\end{equation}

The memory kernel represents the time-dependent viscoelastic resistance,
to which a spherical particle moving in a stationary Maxwell fluid
is subjected. The relation between the complex viscosity and the complex
viscoelastic resistance is $\eta(\omega)=\frac{1}{6\pi a_{0}}\zeta(\omega)$
where $\zeta(\omega)$ is the one-sided Fourier transformation of
$\zeta(t)$ which is defined as $\zeta(\omega):=\int\zeta(t)e^{-i\omega t}dt$.
Hence, $\eta(\omega)=\frac{\eta_{0}}{1+i\omega\tau}$. Now, Equs.
\eqref{eq:1}, \eqref{eq:3} and \eqref{eq:2}, \eqref{eq:4} can
be coupled to get 
\begin{gather}
0=-(\zeta_{0}+k_{1}\tau)\dot{x}_{1}-\varepsilon\tau k_{2}\dot{x}_{2}-k_{1}x_{1}-\varepsilon k_{2}x_{2}\nonumber \\
+\varepsilon(f_{2}+\tau\dot{f}_{2})+(f_{1}+\tau\dot{f}_{1})\label{eq:5}\\
0=-(\zeta_{0}+k_{2}\tau)\dot{x}_{2}-\varepsilon\tau k_{1}\dot{x}_{1}-k_{2}x_{2}-\varepsilon k_{1}x_{1}\nonumber \\
+\varepsilon(f_{1}+\tau\dot{f}_{1})+(f_{2}+\tau\dot{f}_{2})\label{eq:6}
\end{gather}

and then, Equs. \eqref{eq:5} and \eqref{eq:6} can be Fourier transformed
which then can be written in matrix form as
\begin{equation}
\boldsymbol{0=-}\boldsymbol{A}(\omega)\cdot\boldsymbol{x}(\omega)+\boldsymbol{M}(\omega)\cdot\boldsymbol{f}(\omega)\label{eq:7}
\end{equation}

\[
\boldsymbol{A}(\omega)=\begin{pmatrix}-i\omega\gamma_{1}-k_{1} & -\varepsilon k_{2}(i\omega\tau-1)\\
-\varepsilon k_{1}(i\omega\tau-1) & -i\omega\gamma_{2}-k_{2}
\end{pmatrix}
\]

\[
\boldsymbol{M}(\omega)=(1-i\omega\tau)\begin{pmatrix}1 & \varepsilon\\
\varepsilon & 1
\end{pmatrix}
\]

\[
\boldsymbol{x}(\omega)=\begin{pmatrix}x_{1}(\omega)\\
x_{2}(\omega)
\end{pmatrix},\,\boldsymbol{f}(\omega)=\begin{pmatrix}f_{1}(\omega)\\
f_{2}(\omega)
\end{pmatrix}
\]

where, $\gamma_{i}=\zeta_{0}+k_{i}\tau$. Therefore, Eq. \eqref{eq:7}
can be written as
\begin{equation}
\boldsymbol{x}(\omega)=\boldsymbol{A}^{-1}(\omega)\cdot\boldsymbol{M}(\omega)\cdot\boldsymbol{f}(\omega)=\boldsymbol{\chi}(\omega)\cdot\boldsymbol{f}(\omega)\label{eq:8}
\end{equation}

where, $\boldsymbol{\mathbb{\chi}}(\omega)=\boldsymbol{A}^{-1}(\omega)\cdot\boldsymbol{M}(\omega)$
is the response matrix of the two-particle system. The bold-face notation
has been used to represent both matrices and vectors. The response
matrix is given by
\begin{gather}
\boldsymbol{\mathbb{\chi}}(\omega)=\frac{\left\{ (\alpha\omega^{2}+\nu)+i\omega\beta\right\} (-i\omega\tau+1)}{(\alpha\omega^{2}+\nu)^{2}+(\omega\beta)^{2}}\times\nonumber \\
\begin{pmatrix}k_{2}(1-\varepsilon^{2})-i\omega(\gamma_{2}-\varepsilon^{2}k_{2}\tau) & -\varepsilon i\omega\zeta_{0}\\
-\varepsilon i\omega\zeta_{0} & k_{1}(1-\varepsilon^{2})-i\omega(\gamma_{1}-\varepsilon^{2}k_{1}\tau)
\end{pmatrix}\nonumber \\
=\begin{pmatrix}\chi_{11}(\omega) & \chi_{12}(\omega)\\
\chi_{21}(\omega) & \chi_{22}(\omega)
\end{pmatrix}\label{eq:10}
\end{gather}

where,

\begin{gather*}
\alpha=\varepsilon^{2}\tau^{2}k_{1}k_{2}-\gamma_{1}\gamma_{2}\\
\beta=\gamma_{1}k_{2}+\gamma_{2}k_{1}-2\varepsilon^{2}k_{1}k_{2}\tau\\
\nu=k_{1}k_{2}(1-\varepsilon^{2})
\end{gather*}

$\chi_{ij}(\omega)\,(i\,j=1,\,2)$ are the components of the symmetric
matrix$\boldsymbol{\chi}(\omega)$ ($\chi_{12}(\omega)=\chi_{21}(\omega)$).
From Eq. \eqref{eq:10} one can calculate the amplitude and the phase
of the mutual response function of the system which indicates the
response of one particle due to the application of unit force on the
other. The expressions are given by
\begin{multline}
\bigl|\chi_{12}(\omega)\bigr|=\omega\varepsilon\zeta_{0}\times\\
\frac{\sqrt{\bigl\{(\beta-\nu\tau)-\alpha\tau\omega^{2}\bigr\}^{2}\omega^{2}+\bigl\{\nu+(\alpha+\beta\tau)\omega^{2}\bigr\}^{2}}}{(\alpha\omega^{2}+\nu)^{2}+(\omega\beta)^{2}}\label{eq:amp}
\end{multline}
\begin{equation}
\Phi(\omega)=\tan^{-1}\Biggl[\frac{\nu+(\alpha+\beta\tau)\omega^{2}}{\alpha\tau\omega^{2}+(\nu\tau-\beta)\omega}\Biggr]\label{eq:phs}
\end{equation}

\begin{figure*}[!t]
\includegraphics[scale=0.5]{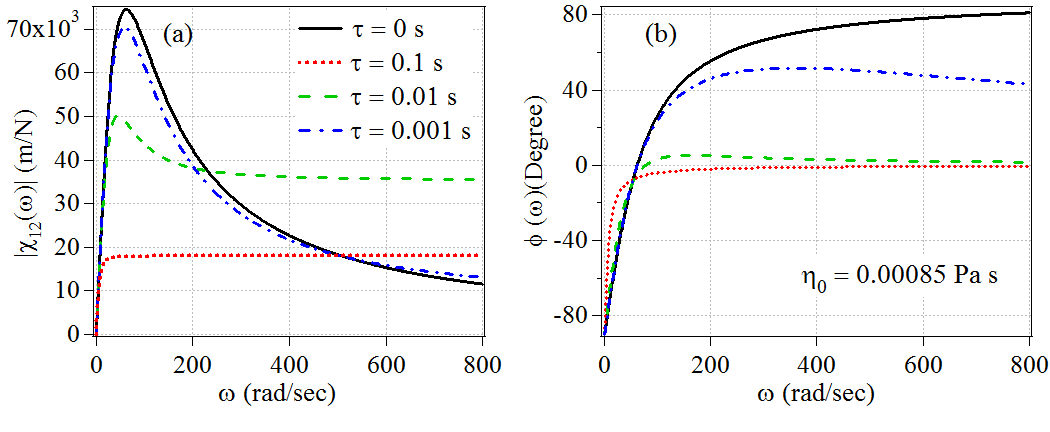}

\caption{The plot of (a) amplitude $|\chi_{12}(\omega)|$ and (b) phase $\phi(\omega)$
of the mutual response function $\chi_{12}(\omega)$ as a function
of the angular frequency $\omega$ for different values of $\tau$.
Black line, blue dot-dashes, green dashes and red dots are for $\tau=0$
s, $0.001$ s, $0.01$ s and $0.1$ s respectively. The zero frequency
viscosity is kept constant at $\eta_{0}=0.00085$ Pa s. Other parameters
have been taken as $a_{0}=1\,\mu m$, $d=10a_{0}$, $k_{1}=k_{2}=1\,\mu N/m$
which are experimentally relevant. }
\label{fig:1}
\end{figure*}

In Fig. \ref{fig:1} the amplitude $\bigl|\chi_{12}(\omega)\bigr|$
and the phase $\Phi(\omega)$ of the mutual response function $\chi_{12}(\omega)$
have been plotted w.r.t angular frequency $\omega$. At $\tau=0$,
the medium is purely viscous and at a particular frequency of the
external drive, the probe particle absorbs maximum energy and the
response is thus maximum. At that frequency, the phase is zero \cite{paul2017direct}.
This frequency in viscous medium is viscosity dependent and thus it
can be used to measure the viscosity of the surrounding medium \cite{paul2018two}.
But if $\tau>0$, the nature of the coupling changes entirely. The
peak in the amplitude decreases with the increase of $\tau$ and after
certain value of $\tau$ the peak vanishes. It is also clear from
the figure that the peak frequency decreases with the increase of
$\tau$ but on the other hand the zero-crossing frequency in the phase
plot increases. The plots can be described from Equs. \eqref{eq:amp}
and \eqref{eq:phs} which are very much different from the reported
functions in a viscous medium.

\subsection{Calculation for the noise correlation matrix:}

It can be assumed that the perturbation $\boldsymbol{f}(\omega)$
in Eq. \eqref{eq:8} on the system is due to the random thermal motions
of the molecules of the surrounding fluid. The random perturbation
(noise) is the manifestation of a large number of equally strong,
independent impulses which change direction rapidly. Therefore, according
to the central limit theorem, the distribution of the noise will be
Gaussian with zero mean ($\langle\boldsymbol{f}(\omega)\rangle=\boldsymbol{0}$).
Now, the inherent elasticity of the fluid enables the system to store
energy and thus the noise becomes correlated. Here, the attempt is
to find out the correlation. In equilibrium, the well known fluctuation-dissipation
theorem (FDT) relates the correlation matrix $\langle\boldsymbol{x}(\omega)\boldsymbol{x}^{\dagger}(\omega)\rangle$
of the system to the deterministic response matrix in the following
form \cite{kubo1966fluctuation}
\begin{align}
\langle\boldsymbol{x}(\omega)\boldsymbol{x}^{\dagger}(\omega)\rangle & =\frac{2k_{B}T}{\omega}\Im[\boldsymbol{\mathbb{\chi}}(\omega)]\label{eq:11-1}\\
\nonumber 
\end{align}

where, $\Im[\boldsymbol{\mathbb{\chi}}(\omega)]$ is the imaginary
part of the response function. The position correlation matrix of
the system of particles can be written as
\begin{equation}
\langle\boldsymbol{x}(\omega)\boldsymbol{x}^{\dagger}(\omega)\rangle=\mathbb{\boldsymbol{\chi}}(\omega)\cdot\langle\boldsymbol{f}(\omega)\boldsymbol{f}^{\dagger}(\omega)\rangle\cdot\mathbb{\boldsymbol{\chi}}^{\dagger}(\omega)\label{eq:16}
\end{equation}

where, Eq. \eqref{eq:8} has been used. Further, the imaginary part
of the response function can be written as
\begin{equation}
\Im[\boldsymbol{\mathbb{\chi}}(\omega)]=\frac{1}{2i}(\boldsymbol{\mathbb{\chi}}(\omega)-\boldsymbol{\mathbb{\chi}}^{*}(\omega))=\frac{1}{2i}(\mathbb{\boldsymbol{\chi}}(\omega)-\mathbb{\boldsymbol{\chi}}^{\dagger}(\omega))\label{eq:17}
\end{equation}

Since, $\boldsymbol{\mathbb{\chi}}(\omega)$ is symmetric so $\boldsymbol{\chi}^{*}(\omega)=\mathbb{\boldsymbol{\chi}}^{\dagger}(\omega)$.
Hence, one can write using Equs. \eqref{eq:11-1}, \eqref{eq:16}
and \eqref{eq:17}
\begin{equation}
\mathbb{\boldsymbol{\chi}}(\omega)\cdot\langle\boldsymbol{f}(\omega)\boldsymbol{f}^{\dagger}(\omega)\rangle\cdot\mathbb{\boldsymbol{\chi}}^{\dagger}(\omega)=\frac{2k_{B}T}{2i\omega}(\boldsymbol{\mathbb{\chi}}(\omega)-\mathbb{\boldsymbol{\chi}}^{\dagger}(\omega))\label{eq:18}
\end{equation}

Thus, 
\begin{eqnarray}
\langle\boldsymbol{f}(\omega)\boldsymbol{f}^{\dagger}(\omega)\rangle & = & \frac{k_{B}T}{i\omega}\mathbb{\boldsymbol{\chi}}^{-1}(\omega)\cdot(\mathbb{\boldsymbol{\chi}}(\omega)-\mathbb{\boldsymbol{\chi}}^{\dagger}(\omega))\cdot(\boldsymbol{\chi}^{\dagger}(\omega))^{-1}\nonumber \\
 & = & \frac{k_{B}T}{i\omega}\biggl((\boldsymbol{\chi}^{\dagger}(\omega))^{-1}-\mathbb{\boldsymbol{\chi}}^{-1}(\omega)\biggr)\label{eq:19}
\end{eqnarray}

Now, the response function can be written as
\[
\mathbb{\boldsymbol{\chi}}(\omega)=(1-i\omega\tau)\boldsymbol{A}^{-1}(\omega)\cdot\boldsymbol{M}_{1}
\]

where, $\boldsymbol{M}(\omega)=(1-i\omega\tau)\boldsymbol{M}_{1}$.
Again,
\[
\chi(\omega)=\frac{(1-i\omega\tau)}{\det(\boldsymbol{A}(\omega))}\widetilde{\mathbb{\boldsymbol{\chi}}}(\omega)=\frac{(1-i\omega\tau)}{\det(\boldsymbol{A}(\omega))}\begin{pmatrix}\widetilde{\chi}_{11}(\omega) & \widetilde{\chi}_{12}(\omega)\\
\widetilde{\chi}_{21}(\omega) & \widetilde{\chi}_{22}(\omega)
\end{pmatrix}
\]

where, $\widetilde{\mathbb{\boldsymbol{\chi}}}(\omega)=adj(\boldsymbol{A}(\omega))\cdot\boldsymbol{M}_{1}$
and the corresponding components are given by
\begin{gather}
\widetilde{\chi}_{11}(\omega)=k_{2}(1-\varepsilon^{2})+i\omega(\varepsilon^{2}k_{2}\tau-\gamma_{2})\label{eq:20}\\
\widetilde{\chi}_{22}(\omega)=k_{1}(1-\varepsilon^{2})+i\omega(\varepsilon^{2}k_{1}\tau-\gamma_{1})\label{eq:21}\\
\widetilde{\chi}_{12}(\omega)=\widetilde{\chi}_{21}(\omega)=-i\omega\varepsilon\zeta_{0}\label{eq:22}
\end{gather}

Therefore,
\begin{gather}
\boldsymbol{\mathbb{\chi}}^{-1}(\omega)=\frac{\det(\boldsymbol{A}(\omega))}{(1-i\omega\tau)}\times\nonumber \\
\frac{1}{\det(\boldsymbol{A}(\omega))\det(\boldsymbol{M}_{1})}\begin{pmatrix}\widetilde{\chi}_{22}(\omega) & -\widetilde{\chi}_{12}(\omega)\\
-\widetilde{\chi}_{21}(\omega) & \widetilde{\chi}_{11}(\omega)
\end{pmatrix}\\
\boldsymbol{\mathbb{\chi}}^{-1}(\omega)=\frac{1}{(1-i\omega\tau)(1-\varepsilon^{2})}\begin{pmatrix}\widetilde{\chi}_{22}(\omega) & -\widetilde{\chi}_{12}(\omega)\\
-\widetilde{\chi}_{21}(\omega) & \widetilde{\chi}_{11}(\omega)
\end{pmatrix}\label{eq:23}
\end{gather}

Similarly,
\begin{equation}
(\mathbb{\boldsymbol{\chi}}^{\dagger}(\omega))^{-1}=\frac{1}{(1+i\omega\tau)(1-\varepsilon^{2})}\begin{pmatrix}\widetilde{\chi}_{22}^{*}(\omega) & -\widetilde{\chi}_{12}^{*}(\omega)\\
-\widetilde{\chi}_{21}^{*}(\omega) & \widetilde{\chi}_{11}^{*}(\omega)
\end{pmatrix}\label{eq:24}
\end{equation}

Using above Equs. \eqref{eq:19}-\eqref{eq:24} one can obtain the
correlation
\begin{equation}
\langle\boldsymbol{f}(\omega)\boldsymbol{f}^{\dagger}(\omega)\rangle=\frac{2k_{B}T\zeta_{0}}{1+(\omega\tau)^{2}}\begin{pmatrix}1 & \varepsilon\\
\varepsilon & 1
\end{pmatrix}^{-1}\label{eq:25}
\end{equation}

The corresponding correlation matrix in time domain is a result of
the Fourier transform of Eq. \eqref{eq:25} and is given by
\begin{equation}
\langle\boldsymbol{f}(t)\boldsymbol{f}^{\dagger}(0)\rangle=\frac{k_{B}T\zeta_{0}}{\tau}e^{-|t|/\tau}\begin{pmatrix}1 & \varepsilon\\
\varepsilon & 1
\end{pmatrix}^{-1}
\end{equation}

This means, the random forces $\boldsymbol{f}(t)$ acting on the system
of particles in Maxwell fluid are exponentially correlated and thus
Markovian. As $\tau$ approaches zero, the correlation converges to
the familiar form in a viscous medium which is given by
\begin{equation}
\langle\boldsymbol{f}(t)\boldsymbol{f}^{\dagger}(0)\rangle=2k_{B}T\zeta_{0}\delta(t)\begin{pmatrix}1 & \varepsilon\\
\varepsilon & 1
\end{pmatrix}^{-1}
\end{equation}

The Markovian random forces can be represented as the solution of
a stochastic differential equation
\begin{equation}
\tau\frac{d}{dt}\boldsymbol{f}(t)+\boldsymbol{f}(t)=\boldsymbol{\xi}(t)\label{eq:30-1}
\end{equation}

where $\boldsymbol{\xi}(t)$ are random forces with correlation
\begin{equation}
\langle\boldsymbol{\xi}(t)\boldsymbol{\xi}^{\dagger}(0)\rangle=2k_{B}T\zeta_{0}\delta(t)\begin{pmatrix}1 & \varepsilon\\
\varepsilon & 1
\end{pmatrix}^{-1}\label{eq:31-1}
\end{equation}

\subsection{Calculation for the correlations and the mean-square displacements
of the particles:\label{subsec:Calculation-for-correlations}}

From Eq. \eqref{eq:30-1}, it can be obtained that the generalized
random force $\boldsymbol{\xi}(t)$ is related to the Markovian random
force $\boldsymbol{f}(t)$ by the relation
\begin{equation}
\boldsymbol{\xi}(\omega)=(1-i\omega\tau)\boldsymbol{f}(\omega)\label{eq:32-1}
\end{equation}

Eq. \eqref{eq:32-1} can be substituted into Eq. \eqref{eq:8} and
one can obtain,
\[
\boldsymbol{x}(\omega)=\frac{1}{(1-i\omega\tau)}\boldsymbol{\chi}(\omega)\cdot\boldsymbol{\xi}(\omega)=\boldsymbol{\chi}_{g}(\omega)\cdot\boldsymbol{\xi}(\omega)
\]

Hence, the response $\boldsymbol{x}(\omega)$ and the generalized
random force is related linearly by the generalized susceptibility
$\boldsymbol{\chi}_{g}(\omega)=\frac{1}{(1-i\omega\tau)}\boldsymbol{\chi}(\omega)$
of the system. Thus, in equilibrium, the position correlation matrix
of the system due to the thermal motions of the particles can be obtained
in terms of the generalized susceptibility $\boldsymbol{\chi}_{g}(\omega)$
as
\begin{align}
\boldsymbol{C}(\omega) & =\frac{2k_{B}T}{\omega}\Im[\boldsymbol{\mathbb{\chi}}_{g}(\omega)]\label{eq:11}\\
\begin{pmatrix}C_{11}(\omega) & C_{12}(\omega)\\
C_{21}(\omega) & C_{22}(\omega)
\end{pmatrix} & =\frac{2k_{B}T}{\omega}\Im\Biggl[\frac{1}{(1-i\omega\tau)}\begin{pmatrix}\chi_{11}(\omega) & \chi_{12}(\omega)\\
\chi_{21}(\omega) & \chi_{22}(\omega)
\end{pmatrix}\Biggr]\label{eq:12}
\end{align}

$\boldsymbol{C}(\omega)$ is the correlation matrix in frequency domain
and $C_{ij}(\omega)\,(i\,j=1,\,2)$ are the corresponding components,
$k_{B}$ is the Boltzmann constant and $T$ is the temperature. Now,
from Equs. \eqref{eq:12} and \eqref{eq:10} one can get
\begin{gather}
C_{11}(\omega)=\frac{2k_{B}T}{(\alpha\omega^{2}+\nu)^{2}+(\omega\beta)^{2}}\times\nonumber \\
\biggl[\bigl\{\alpha(\varepsilon^{2}k_{2}\tau-\gamma_{2})\bigr\}\omega^{2}+\nonumber \\
\bigl\{\beta k_{2}(1-\varepsilon^{2})+\nu(\varepsilon^{2}k_{2}\tau-\gamma_{2})\bigr\}\biggr]\label{eq:13}
\end{gather}

\begin{gather}
C_{22}(\omega)=\frac{2k_{B}T}{(\alpha\omega^{2}+\nu)^{2}+(\omega\beta)^{2}}\times\nonumber \\
\biggl[\bigl\{\alpha(\varepsilon^{2}k_{1}\tau-\gamma_{1})\bigr\}\omega^{2}+\nonumber \\
\bigl\{\beta k_{1}(1-\varepsilon^{2})+\nu(\varepsilon^{2}k_{1}\tau-\gamma_{1})\bigr\}\biggr]\label{eq:14}
\end{gather}

\begin{gather}
C_{12}(\omega)=C_{21}(\omega)=-\frac{2k_{B}T}{(\alpha\omega^{2}+\nu)^{2}+(\omega\beta)^{2}}\times\nonumber \\
\biggl[\varepsilon\zeta_{0}\bigl\{\alpha\omega^{2}+\nu\bigr\}\biggr]\label{eq:15}
\end{gather}

Now, the position correlation functions of the particles in the time
domain can be obtained by Fourier transforming Equs. \eqref{eq:13},
\eqref{eq:14} and \eqref{eq:15}. Which yields,

\begin{gather}
C_{ii}(t)=\left<x_{i}(t)x_{i}(0)\right>=\nonumber \\
\frac{1}{\nu_{1}}\Biggl[\frac{a_{ii}-\frac{b_{ii}}{4}(c-\nu_{1})^{2}}{\biggl\{\left(\frac{c-\nu_{1}}{2}\right)^{2}+\left(\frac{c-\nu_{1}}{2}\right)c+\omega_{0}\biggr\}}\exp\biggl(-\left(\frac{c-\nu_{1}}{2}\right)t\biggr)\nonumber \\
+\frac{\frac{b_{ii}}{4}(c+\nu_{1})^{2}-a_{ii}}{\biggl\{\left(\frac{c+\nu_{1}}{2}\right)^{2}+\left(\frac{c+\nu_{1}}{2}\right)c+\omega_{0}\biggr\}}\exp\biggl(-\left(\frac{c+\nu_{1}}{2}\right)t\biggr)\Biggr]\label{eq:26}
\end{gather}

\begin{gather}
C_{ij}(t)=\left<x_{i}(t)x_{j}(0)\right>=\nonumber \\
\frac{1}{\nu_{1}}\Biggl[\frac{a_{ij}-\frac{b_{ij}}{4}(c-\nu_{1})^{2}}{\biggl\{\left(\frac{c-\nu_{1}}{2}\right)^{2}+\left(\frac{c-\nu_{1}}{2}\right)c+\omega_{0}\biggr\}}\exp\biggl(-\left(\frac{c-\nu_{1}}{2}\right)t\biggr)\nonumber \\
+\frac{\frac{b_{ij}}{4}(c+\nu_{1})^{2}-a_{ij}}{\biggl\{\left(\frac{c+\nu_{1}}{2}\right)^{2}+\left(\frac{c+\nu_{1}}{2}\right)c+\omega_{0}\biggr\}}\exp\biggl(-\left(\frac{c+\nu_{1}}{2}\right)t\biggr)\Biggr]\label{eq:27}
\end{gather}

\begin{equation}
C_{ij}(t)=C_{ji}(t)\label{eq:28}
\end{equation}

where, $a_{ii}=\frac{2k_{B}T}{\alpha^{2}}\bigl\{\beta k_{j}(1-\varepsilon^{2})+\nu(\varepsilon^{2}k_{j}\tau-\gamma_{j})\bigr\}$,
$b_{ii}=\frac{2k_{B}T}{\alpha^{2}}\bigl\{\alpha(\varepsilon^{2}k_{j}\tau-\gamma_{j})\bigr\}$,
$a_{ij}=-\frac{2k_{B}T}{\alpha^{2}}\varepsilon\zeta_{0}\nu$, $b_{ij}=-\frac{2k_{B}T}{\alpha}\varepsilon\zeta_{0}$
$\omega_{0}=-\frac{\nu}{\alpha}$, $c=-\frac{\beta}{\alpha}$ and
$\nu_{1}=\sqrt{c^{2}-4\omega_{0}}$. $i,\,j=1,\,2;\,i\neq j$. The
mean-square displacement functions (MSD) is related to the correlation
functions as
\begin{gather}
\left\langle \Delta x_{i}^{2}(t)\right\rangle =2\left[\left\langle x_{i}^{2}(0)\right\rangle -\left\langle x_{i}(t)x_{i}(0)\right\rangle \right]\nonumber \\
=\frac{2}{\nu_{1}}\Biggl[\frac{a_{ii}-\frac{b_{ii}}{4}(c-\nu_{1})^{2}}{\biggl\{\left(\frac{c-\nu_{1}}{2}\right)^{2}+\left(\frac{c-\nu_{1}}{2}\right)c+\omega_{0}\biggr\}}\Biggl(1-\exp\biggl(-\left(\frac{c-\nu_{1}}{2}\right)t\biggr)\Biggr)\nonumber \\
+\frac{\frac{b_{ii}}{4}(c+\nu_{1})^{2}-a_{ii}}{\biggl\{\left(\frac{c+\nu_{1}}{2}\right)^{2}+\left(\frac{c+\nu_{1}}{2}\right)c+\omega_{0}\biggr\}}\Biggl(1-\exp\biggl(-\left(\frac{c+\nu_{1}}{2}\right)t\biggr)\Biggr)\Biggr]\label{eq:29}
\end{gather}

\begin{figure}[!tp]
\includegraphics[scale=0.4]{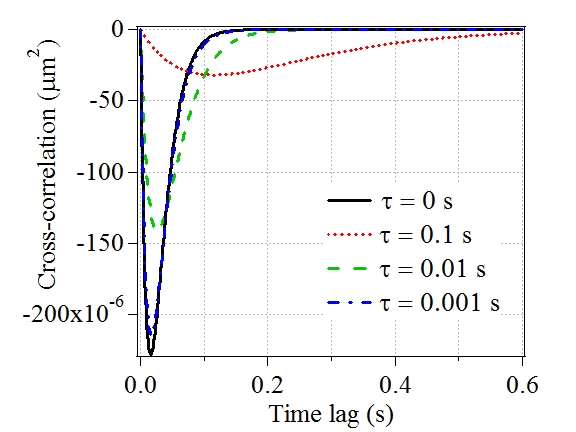}

\caption{The plot of the cross-correlation function with respect to time-lag.
The parameters are same as described in Fig. \ref{fig:1}. Black line,
blue dot-dashes, green dashes and red dots represent $\tau=0$ s,
$0.001$ s, $0.01$ s and $0.1$ s respectively.}
\label{fig:2}

\end{figure}

\begin{figure}[!tp]
\includegraphics[scale=0.4]{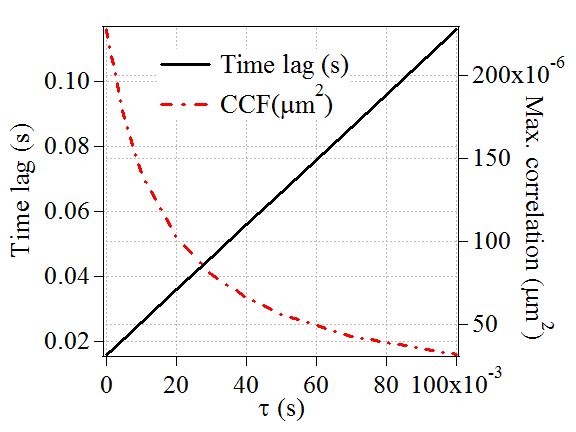}\caption{The plot of the maximum correlations and the corresponding time lags
with respect to the Maxwell time constant $\tau$. Other parameters
are chosen as in Fig. \ref{fig:1}. }

\label{fig:4}
\end{figure}

In Fig. \ref{fig:2}, the cross-correlation functions for different
$\tau$ has been plotted. With the increase of $\tau$, the maximum
correlation between the particles appear in larger time-lag which
increases linearly with $\tau$ and the corresponding correlation
decreases exponentially. It has been shown in Fig. \ref{fig:4}.

\subsection{Coupled motion in a viscous fluid:}

The coupled dynamics in viscous fluid can be obtained by assuming
$\tau\rightarrow0$ in the above equations. It is clear from Equs.
\eqref{eq:13}, \eqref{eq:14} and \eqref{eq:15} that the correlation
functions in frequency domain converge to
\begin{equation}
C_{ii}(\omega)=2k_{B}T\frac{\bigl[\frac{\omega^{2}}{\zeta_{0}}+\frac{k_{j}^{2}}{\zeta_{0}^{3}}(1-\varepsilon^{2})\bigr]}{\bigr[\bigl\{\omega^{2}-\frac{k_{i}k_{j}}{\zeta_{0}^{2}}(1-\varepsilon^{2})\bigl\}^{2}+\bigl\{(\frac{k_{i}+k_{j}}{\zeta_{0}}\omega)^{2}\bigr\}\bigl]}\label{eq:30}
\end{equation}

and 
\begin{gather}
C_{ij}(\omega)=C_{ji}(\omega)=\nonumber \\
2k_{B}T\frac{\varepsilon\bigl[\frac{k_{i}k_{j}}{\zeta_{0}^{3}}(1-\varepsilon^{2})-\frac{\omega^{2}}{\zeta_{0}})\bigr]}{\bigr[\bigl\{\omega^{2}-\frac{k_{i}k_{j}}{\zeta_{0}^{2}}(1-\varepsilon^{2})\bigl\}^{2}+\bigl\{(\frac{k_{i}+k_{j}}{\zeta_{0}}\omega)^{2}\bigr\}\bigl]}\label{eq:31}
\end{gather}

$\varepsilon\rightarrow0$ implies zero hydrodynamic coupling which
yields 
\begin{equation}
C_{ii}(\omega)=\frac{2k_{B}T/\zeta_{0}}{\omega^{2}+\frac{k_{i}^{2}}{\zeta_{0}^{2}}}\label{eq:33}
\end{equation}

and 
\begin{equation}
C_{ij}(\omega)=C_{ji}(\omega)=0\label{eq:34}
\end{equation}

\begin{figure}[!tp]
\includegraphics[scale=0.4]{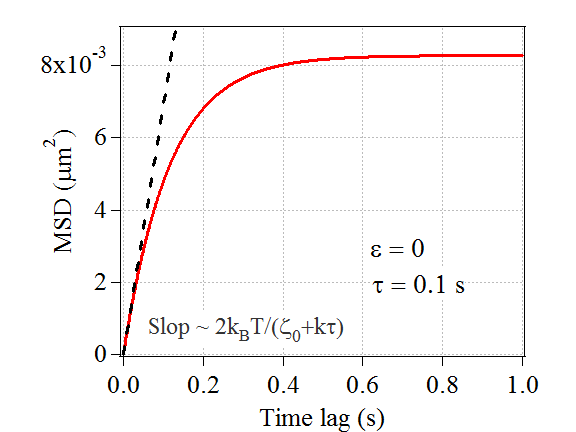}

\caption{The uncoupled mean-square displacement (MSD) of one of the particle
against time-lag for $\tau=0.1\,s$ in red line. $k=1\,\mu N/m$.
The black dashed line is the straight line fit to the initial portion
of the MSD curve. The slope of the straight line is $\frac{2k_{B}T}{\zeta_{0}+k_{i}\tau}$.}
\label{fig:3}
\end{figure}

In the similar way as described in the subsection \ref{subsec:Calculation-for-correlations},
the correlation functions in a viscous fluid in the time domain are
of similar form as Equs. \eqref{eq:26}, \eqref{eq:27} and \eqref{eq:28}
where the parameters will be changed to $a_{ii}=\frac{2k_{B}T}{^{\zeta_{0}^{3}}}\bigl\{ k_{j}^{2}(1-\varepsilon^{2})\bigr\}$,
$b_{ii}=\frac{2k_{B}T}{\zeta_{0}}$, $a_{ij}=-\frac{2k_{B}T}{\zeta_{0}^{3}}\varepsilon\bigl\{ k_{i}k_{j}(1-\varepsilon^{2})\bigr\}$,
$b_{ij}=-\frac{2k_{B}T}{\zeta_{0}}\varepsilon$ $\omega_{0}=\frac{k_{i}k_{j}}{\zeta_{0}^{2}}(1-\varepsilon^{2})$,
$c=\frac{k_{i}+k_{j}}{\zeta_{0}}$ and $\nu_{1}=\sqrt{c^{2}-4\omega_{0}}$.
Where, $i,\,j=1,\,2$ and $i\neq j$. In the expression of MSD, Eq.
\eqref{eq:29}, one can put $\tau\rightarrow0$ and $k_{i}\rightarrow0$
and can get $\left\langle \Delta x_{i}^{2}(t)\right\rangle =\frac{2k_{B}T}{\zeta_{0}}t=2Dt.$
$D$ is the diffusion constant. For uncoupled motion in viscoelastic
fluid, $\varepsilon=0$ and $\tau\neq0$ and then $\left\langle \Delta x_{i}^{2}(t)\right\rangle =\frac{2k_{B}T}{\zeta_{0}+k_{i}\tau}t=2D_{r}t$.
$D_{r}=\frac{k_{B}T}{\zeta_{0}+k_{i}\tau}$ is the generalized diffusion
coefficient \cite{volkov1996non}. It has been shown in Fig. \ref{fig:3}
as a linear fit to the MSD values corresponding to very low time-lags
where the effect of the trap is negligible. 

\section{Conclusions}

In conclusion, a phenomenological theory of the equilibrium dynamics
of two hydrodynamically coupled Brownian harmonic oscillators in a
Maxwell fluid in low Reynolds numbers approximation has been presented.
The response functions have been calculated and shown that these are
drastically different from the reported functions in a viscous fluid.
For instance, the dependency of the mutual response function on the
Maxwell time constant $\tau$ which has been shown. Therefore, the
formulated response functions derived in this paper can be used to
perform rheological measurements in Maxwell fluid as it is done before
in a viscous fluid. Further, the correlation between the noises present
on the particles has been calculated and shown that such problem of
the coupled Brownian motion with the simplest viscoelastic liquid
can be reduced to the statistical description of an extended dynamical
system subjected to a delta-correlated random force. Consequently,
the generalized susceptibility of the system has been calculated and
then used to calculate the position correlation functions in the frequency
and the time domain. It is clear from the cross-correlation function
that the two particles have time-delayed correlation and the time
delay is a linear function of $\tau$ . In addition, the corresponding
correlation depends on $\tau$ exponentially. Thereupon, the mean-square
displacement functions of the two particles have been calculated which
reveals the generalized diffusion coefficient $D_{r}=\frac{2k_{B}T}{\zeta_{0}+k_{i}\tau}$
in a Maxwell fluid in the approximation of the negligible hydrodynamic
coupling, which is known to scientific community. The statistical
descriptions which are derived in this paper, converge to the description
in a purely viscous fluid when the Maxwell time constant $\tau$ tends
to zero . Only the zero-frequency viscosity $\eta_{0}$ is incorporated
in the Maxwell model as dissipation mechanism. Thus, the back ground
viscosity, which is defined as the viscosity of a viscoelastic medium
at $\omega\rightarrow\infty$, is neglected. Hence, in future, the
reported theory can be extended using more generalized forms, like
the jeffreys' model, representing viscoelasticity which can disclose
much more interesting facts.
\begin{acknowledgments}
The author wants to acknowledge Dr Ayan Banerjee, Associate Professor
at Indian Institute of Science Education and Research, Kolkata for
his wise advice and guidance as a PhD mentor. The author would like
to thank Mrs Puspa Saha for her help in the calculations and Mr Sudipta
Saha for his important suggestions. The author also would like to
thank the Indian Institute of Science Education and Research, Kolkata
for providing the Senior research fellowship to the author. 
\end{acknowledgments}

\bibliographystyle{apsrev4-1}
\addcontentsline{toc}{section}{\refname}\bibliography{Maxwell}

\end{document}